\documentclass[10pt,a4paper,twoside]{article}
\usepackage{epsfig}
\usepackage{baltlat6}
\usepackage{array}
\usepackage{here}
\begin{document}
\ \ \vspace{0.5mm} \setcounter{page}{1} \vspace{8mm}

\titlehead{Baltic Astronomy, vol.20, 587-590, 2011}

\titleb{CROSS SECTIONS FOR ELECTRON IMPACT EXCITATION OF O VI LINES}

\begin{authorl}
\authorb{Haykel Elabidi}{1},
\authorb{Sylvie Sahal-Br\'{e}chot}{2} and
\authorb{N\'{e}bil Ben Nessib}{3}
\end{authorl}

\begin{addressl}
\addressb{1}{Groupe de Recherche en Physique Atomique et Astrophysique,
Facult\'{e} des Sciences de Bizerte, Universit\'{e} de Carthage,
Tunisia; haykel.elabidi@fsb.rnu.tn}
\addressb{2}{LERMA, Observatoire de Paris, CNRS, Universit\'{e} Pierre et Marie Curie, \\
Place Jules Janssen, 92190 Meudon, France; sylvie.sahal-brechot@obspm.fr}
\addressb{3}{Groupe de Recherche en Physique Atomique et Astrophysique,
INSAT, Universit\'{e} de Carthage, Tunisia;
nebil.bennessib@planet.tn}
\end{addressl}

\submitb{Received: 2011 August 08; accepted: 2011 August 15}

\begin{summary}
Radiative atomic data and electron impact excitation cross
sections for the $2s-2p$ transitions in O VI for transitions among
the fine structure levels belonging to the $1s^{2}nl$ ($2\leq
n\leq5$) configurations have been calculated. We have extended the
calculations of fine structure collision strengths up to 140 Ry
and have compared our results at energies below 63 Ry to the
R-matrix ones.
\end{summary}

\begin{keywords}
Impact excitation by electrons, cross sections, distorted wave
method, plasma diagnostics, oscillator strengths
\end{keywords}

\resthead{Electron impact excitation cross sections of O VI lines}
{H. Elabidi \emph{et al.}}

\sectionb{1}{INTRODUCTION}
Transitions in O VI have been observed in stellar spectra, in
white dwarfs, in the solar corona, and in the solar transition
region, where the two resonance lines at $1031.924$ $\rm \AA$ and
$1037.614$ $\rm \AA$ are among the brightest emitted (Lozano et
al. 2001). Excitation cross sections for this ion are very
important for the spectroscopic diagnostics. We present in this
paper, energy levels, oscillator strengths, electron impact
collision strengths and cross sections for the O VI ion. Atomic
data are compared to NIST (physics.nist.gov) and to Aggarwal \&
Keenan (2004) results. Cross sections for energies near the
excitation threshold are compared to the experimental results of
Lozano et al. (2001). Collision strengths are compared to results
of Aggarwal \& Keenan (2004) at energies up to 63 Ry. We extend
our calculations up to 140 Ry.

\sectionb{2}{COMPUTATIONAL PROCEDURE}
The atomic structure has been computed using the UCL (University
College, London) computer package SUPERSTRUCTURE (SST) of Eissner
et al. (1974). This code takes into account configuration
interaction. Relativistic corrections (spin-orbit, mass, Darwin
and one-body) are introduced according to the Breit-Pauli. The
electron scattering calculation has been performed in the
distorted wave approximation using the DISTORTED WAVE code
(Eissner 1998). Fine structure collision parameters have been
obtained by the JAJOM code (Saraph 1978) for low partial wave $l$
of the incoming electron ($l$=29). This code transforms the
transition matrix elements from $LS$ into $LSJ$ coupling using
Term Coupling Coefficients given by the SST code. Contributions to
collision strengths for 30$\leq l\leq 50$ have been taken into
account through the Coulomb-Bethe approximation for the dipole
transitions and a geometric series for the non-dipole ones.

\sectionb{3}{RESULTS AND DISCUSSIONS}

Energies of the 24 fine structure levels belonging to the
$1s^{2}nl$ ($2\leq n\leq5$) configurations and oscillator
strengths of some allowed transitions are presented in Table 1.
Comparison with NIST results and with those of Aggarwal \& Keenan
(2004) where the authors used the fully relativistic GRASP code of
Dally et al. (1989) gives an agreement better than 1\% for level
energies and does not exceed 2\% for oscillator strengths.

\begin{table}[!tH]
\begin{center}
\vbox{\footnotesize\tabcolsep=3pt
\parbox[c]{124mm}{\baselineskip=10pt
{\smallbf\ \ Table 1.}{\small\ O VI energy levels and oscillator
strengths of some allowed transitions compared to NIST and GRASP
results.\lstrut}}
\begin{tabular}{ccccccccccc}
\hline \noalign{\smallskip}\multicolumn{3}{c}{Level designation} &
\multicolumn{3}{c}{Energy (Ry)} &  &
Transition & \multicolumn{3}{c}{$f_{ij}$} \\
\noalign{\smallskip}\hline \noalign{\smallskip} $i$ & Conf. &
level
& Present & NIST & GRASP &  & $i-j$ & Present & NIST & GRASP \\
\noalign{\smallskip}\hline
$1$ & $1s^{2}2s$ & $^{2}S_{1/2}$&$0.00000$ & $0.0000$ & $0.00000$ &  & $1-2$ & $0.0667$ & $0.0661$ & $0.0673$ \\
$2$ & $1s^{2}2p$ & $^{2}P_{1/2}^{\mathrm{o}}$ & $0.88049$ &$0.8782$ & $0.88628$ &  & $1-3$ & $0.1342$ & $0.1327$ & $0.1355$ \\
$3$ & $1s^{2}2p$ & $^{2}P_{3/2}^{\mathrm{o}}$ & $0.88522$ &$0.8831$ & $0.89103$ &  & $1-5$ & $0.0893$ & $0.0885$ & $0.0873$ \\
$4$ & $1s^{2}3s$ & $^{2}S_{1/2}$ & $5.82305$ & $5.8325$ &$5.82511$ &  & $1-6$ & $0.1782$ & $0.1770$ & $0.1704$ \\
$5$ & $1s^{2}3p$ & $^{2}P_{1/2}^{\mathrm{o}}$ & $6.05983$ &$6.0701$ & $6.06464$ &  & $1-10$ & $0.0263$ & $0.0247$ & $0.0244$ \\
$6$ & $1s^{2}3p$ & $^{2}P_{3/2}^{\mathrm{o}}$ & $6.06122$ &$6.0715$ & $6.06604$ &  & $1-11$ & $0.0526$ & $0.0494$ & $0.0488$ \\
$7$ & $1s^{2}3d$ & $^{2}D_{3/2}$ & $6.13463$ & $6.1476$ &$6.13912$ &  & $2-4$ & $0.0289$ & $0.0289$ & $0.0287$ \\
$8$ & $1s^{2}3d$ & $^{2}D_{5/2}$ & $6.13505$ & $6.1481$ &$6.13954$ &  & $2-7$ & $0.6590$ & $0.6576$ & $0.6595$ \\
$9$ & $1s^{2}4s$ & $^{2}S_{1/2}$ & $7.75846$ & $7.7703$ &$7.76151$ &  & $2-9$ & $0.0058$ & $0.0057$ & $0.0056$ \\
$10$ & $1s^{2}4p$ & $^{2}P_{1/2}^{\mathrm{o}}$ & $7.85449$ &$7.8673$ & $7.85910$ &  & $3-4$ & $0.0290$ & $0.0290$ & $0.0288$ \\
$11$ & $1s^{2}4p$ & $^{2}P_{3/2}^{\mathrm{o}}$ & $7.85508$ &$7.8679$ & $7.85969$ &  & $3-7$ & $0.0660$ & $0.0656$ & $0.0660$ \\
$12$ & $1s^{2}4d$ & $^{2}D_{3/2}$ & $7.88581$ & $7.8996$ &$7.89029$ &  & $3-8$ & $0.5936$ & $0.5915$ & $0.5940$ \\
$13$ & $1s^{2}4d$ & $^{2}D_{5/2}$ & $7.88599$ & $7.8998$ &$7.89046$ &  & $3-9$ & $0.0058$ & $0.0057$ & $0.0057$ \\
$14$ & $1s^{2}4f$ & $^{2}F_{5/2}^{\mathrm{o}}$ & $7.88710$ &$7.9014$ & $7.89157$ &  & $4-5$ & $0.1106$ & $0.1114$ & $0.1122$ \\
$15$ & $1s^{2}4f$ & $^{2}F_{7/2}^{\mathrm{o}}$ & $7.88719$ &$7.9015$ & $7.89166$ &  & $4-6$ & $0.2226$ & $0.2239$ & $0.2259$ \\
$16$ & $1s^{2}5s$ & $^{2}S_{1/2}$ & $8.63249$ & $8.6451$ &$8.63594$ &  & $4-10$ & $0.0946$ & $0.0922$ & $0.0917$ \\
$17$ & $1s^{2}5p$ & $^{2}P_{1/2}^{\mathrm{o}}$ & $8.68047$ &$8.6942$ & $8.68501$ &  & $4-11$ & $0.1886$ & $0.1849$ & $0.1827$ \\
$18$ & $1s^{2}5p$ & $^{2}P_{3/2}^{\mathrm{o}}$ & $8.68077$ &$8.6942$ & $8.68531$ &  & $5-7$ & $0.0473$ & $0.0492$ & $0.0470$ \\
$19$ & $1s^{2}5d$ & $^{2}D_{3/2}$ & $8.69644$ & $8.7104$ &$8.70091$ &  & $5-9$ & $0.0644$ & $0.0637$ & $0.0641$ \\
$20$ & $1s^{2}5d$ & $^{2}D_{5/2}$ & $8.69653$ & $8.7104$ &$8.70100$ &  & $6-7$ & $0.0046$ & $0.0048$ & $0.0046$ \\
$21$ & $1s^{2}5f$ & $^{2}F_{5/2}^{\mathrm{o}}$ & $8.69716$ &$8.7115$ & $8.70163$ &  & $6-8$ & $0.0420$ & $0.0435$ & $0.0418$ \\
$22$ & $1s^{2}5f$ & $^{2}F_{7/2}^{\mathrm{o}}$ & $8.69721$ &$8.7115$ & $8.70168$ &  & $6-9$ & $0.0647$ & $0.0638$ & $0.0643$ \\
$23$ & $1s^{2}5g$ & $^{2}G_{7/2}$ & $8.69722$ & $8.7116$ &$8.70169$ &  & $7-10$ & $0.0127$ & $0.0127$ & $0.0127$ \\
$24$ & $1s^{2}5g$ & $^{2}G_{9/2}$ & $8.69724$ & $8.7116$&$8.70171$ &  & $7-11$ & $0.0025$ & $0.0025$ & $0.0025$ \\
\hline
\end{tabular}
}
\end{center}
\vskip-6mm
\end{table}

The calculated cross sections for energies near the excitation
threshold of the $2s-2p$ transition are in good agreement with the
experimental results of Lozano et al. (2001) as it is shown in
Figure 1. Figures 2. and 3. display collision strengths for some
transitions, we find that some of them agree well with the DARC
R-matrix calculations, where the authors adopted the Dirac Atomic
R-matrix Code (DARC) of Norrington \& Grant (Private
communication), but for other transitions, they have the same
behavior with energy but they are not in good agreement with the
R-matrix results. Figure 4. shows also that for some other
transitions, present collision strengths are not in agreement with
the R-matrix calculations. In Table 2 are presented our collision
strengths for three energies above thresholds and compared to DARC
results. The two calculations agree within about 24\%, 11\% and
12\% respectively for 15, 45 and 63 Ry. We have extended in the
present work our fine structure collision strengths for electron
energies up to 140 Ry.

Since collision parameters are used in our \emph{ab initio}
calculations of line broadening (Elabidi et al. 2008, 2009,
Elabidi \& Sahal-Br\'{e}chot 2011), comparison with experimental
and other theoretical results of these parameters can be a
powerful tool to check our line broadening calculations.

\begin{figure}[h]
\begin{minipage}{14pc}
\includegraphics[width=16pc]{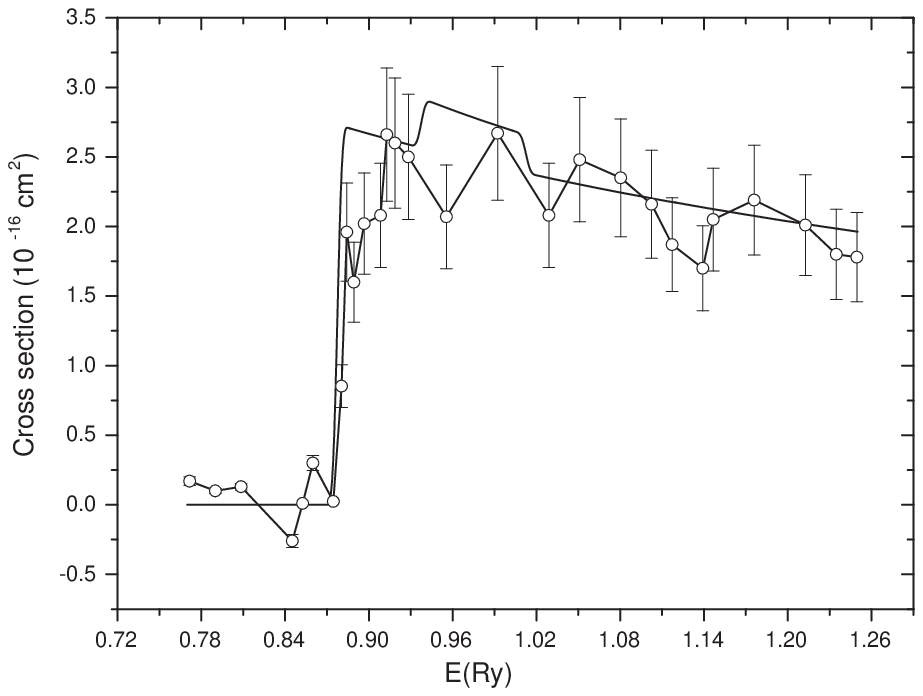}
\captionb{1}{\label{fig3} Present cross sections (solid line) of
the $2s$ $^{2}$S$-2p$ $^{2}$P$^{\rm o}$ transition as a function
of electron energy near the excitation threshold. Experimental
results of Lozano et al. (2001) (solid line $+$ circles)}
\end{minipage}\hspace{2pc}
\begin{minipage}{14pc}
\includegraphics[width=15pc]{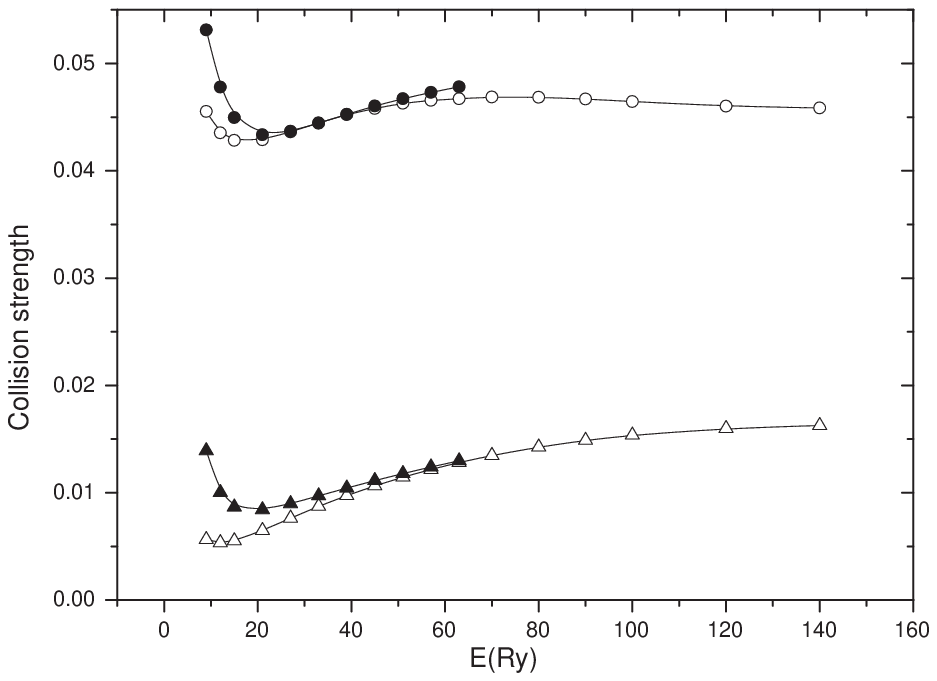}
\captionb{2}{\label{fig2}Present collision strengths (open
symbols) as a function of electron energy compared to R-matrix
DARC results (solid symbols) for the two transitions: $2p$
$^{2}$P$_{1/2}^{\rm o}-3p$ $^{2}$P$_{3/2}^{\rm o}$ (circles) and
$2p$ $^{2}$P$_{1/2}^{\rm o}-4s$ $^{2}$S$_{1/2}$ (triangles).}
\end{minipage}
\end{figure}

\begin{figure}[h]
\begin{minipage}{14pc}
\includegraphics[width=15pc]{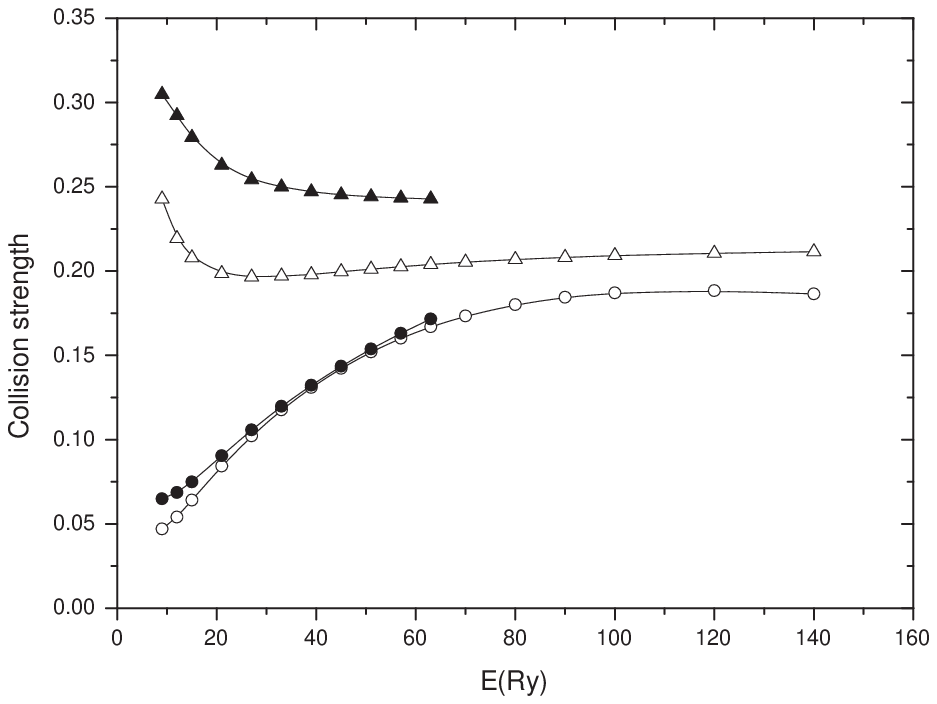}
\captionb{3}{\label{fig3} Present collision strengths (open
symbols) as a function of electron energy compared to R-matrix
DARC results (solid symbols) for the two transitions: $2s$
$^{2}$S$_{1/2}-2p$ $^{2}$P$_{1/2}^{\rm o}$ (circles) and $2s$
$^{2}$S$_{1/2}-2p$ $^{2}$P$_{3/2}^{\rm o}$ (triangles).}
\end{minipage}\hspace{2pc}
\begin{minipage}{14pc}
\includegraphics[width=15pc]{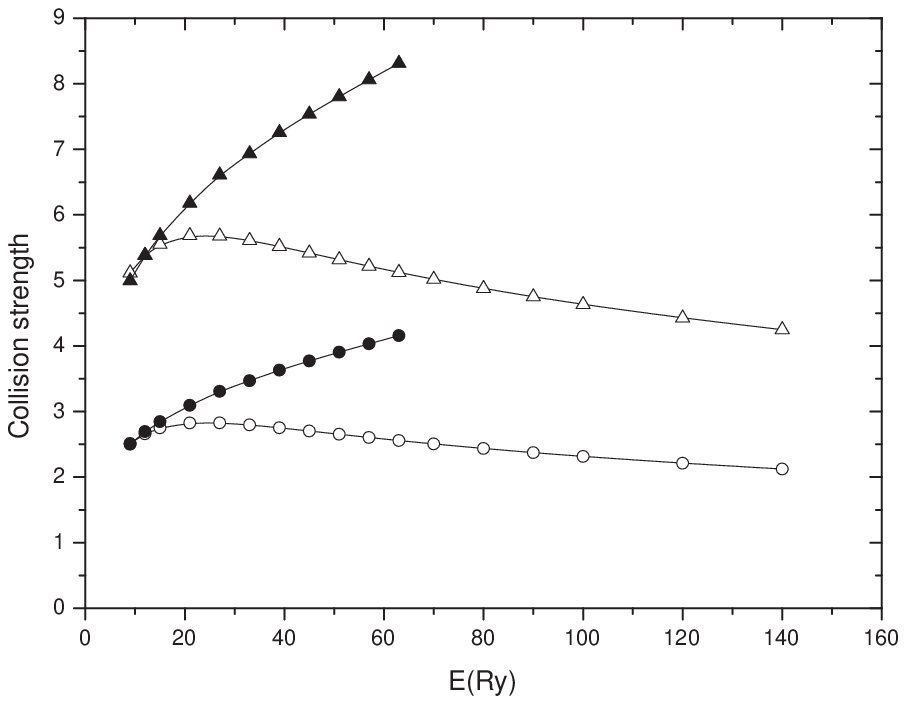}
\captionb{4}{\label{fig2}Present collision strengths (open
symbols) as a function of electron energy compared to R-matrix
DARC results (solid symbols) for the two transitions: $2p$
$^{2}$P$_{3/2}^{\rm o}-3s$ $^{2}$S$_{1/2}$ (circles) and $2p$
$^{2}$P$_{1/2}^{\rm o}-2p$ $^{2}$P$_{3/2}^{\rm o}$ (triangles).}
\end{minipage}
\end{figure}

\begin{table}[!tH]
\begin{center}
\vbox{\footnotesize\tabcolsep=3pt
\parbox[c]{124mm}{\baselineskip=10pt
{\smallbf\ \ Table 2.}{\small\ O VI collisions strengths compared
to R-matrix DARC results.\lstrut}}
\begin{tabular}{cccccccccccc}
\hline
\multicolumn{2}{c}{Transition} &  & \multicolumn{9}{c}{Energy} \\
\cline{1-2}\cline{4-12} &  &  & \multicolumn{2}{c}{$15$ Ry} &  &
\multicolumn{2}{c}{$45$ Ry} &  &
\multicolumn{2}{c}{$63$ Ry} & $\textbf{140}$\textbf{\ Ry} \\
\cline{4-5}\cline{7-8}\cline{10-12} $i$ & $j$ &  & Present & DARC
&  & Present & DARC &  & Present & DARC & \textbf{Present} \\
\hline
$1$ & $2$ &  & $2.751-0$ & $2.846-0$ &  & $2.703-0$ &$3.772-0$ &  & $2.557-0$ & $4.158-0$ & $\mathbf{2.123-0}$\\
$1$ & $3$ &  & $5.552-0$ & $5.685-0$ &  & $5.418-0$ & $7.537-0$ && $5.122-0$ & $8.311-0$ & $\mathbf{4.249-0}$ \\
$1$ & $4$ &  & $1.384-1$ & $1.982-1$ &  & $1.897-1$ & $2.234-1$ && $2.000-1$ & $2.276-1$ & $\mathbf{2.220-1}$ \\
$1$ & $5$ &  & $5.831-2$ & $6.395-2$ &  & $1.525-1$ & $1.492-1$ && $1.850-1$ & $1.824-1$ & $\mathbf{2.278-1}$ \\
$1$ & $6$ &  & $1.175-1$ & $1.272-1$ &  & $3.033-1$ & $2.970-1$ && $3.681-1$ & $3.631-1$ & $\mathbf{4.536-1}$ \\
$1$ & $7$ &  & $1.602-1$ & $1.618-1$ &  & $2.165-1$ & $2.266-1$ && $2.271-1$ & $2.392-1$ & $\mathbf{2.326-1}$ \\
$1$ & $8$ &  & $2.410-1$ & $2.427-1$ &  & $3.257-1$ & $3.399-1$ && $3.415-1$ & $3.587-1$ & $\mathbf{3.493-1}$ \\
$1$ & $9$ &  & $2.563-1$ & $3.833-2$ &  & $3.760-2$ & $4.320-2$ && $4.056-2$ & $4.415-2$ & $\mathbf{4.614-2}$ \\
$1$ & $10$ &  & $1.158-2$ & $1.697-2$ &  & $3.257-2$ & $3.415-2$&& $4.054-2$ & $4.118-2$ & $\mathbf{5.549-2}$ \\
$2$ & $3$ &  & $2.078-1$ & $2.793-1$ &  & $1.996-1$ & $2.452-1$ && $2.039-1$ & $2.428-1$ & $\mathbf{2.115-1}$ \\
$2$ & $4$ &  & $3.173-2$ & $3.725-2$ &  & $7.018-2$ & $7.132-2$ && $8.256-2$ & $8.524-2$ & $\mathbf{9.255-2}$ \\
$2$ & $5$ &  & $1.519-1$ & $2.161-1$ &  & $1.895-1$ & $2.438-1$ && $2.007-1$ & $2.483-1$ & $\mathbf{2.233-1}$ \\
$2$ & $6$ &  & $4.284-2$ & $4.496-2$ &  & $4.581-2$ & $4.604-2$ && $4.672-2$ & $4.783-2$ & $\mathbf{4.585-2}$ \\
$2$ & $7$ &  & $1.131-0$ & $1.159-0$ &  & $2.134-0$ & $2.117-0$ && $2.430-0$ & $2.449-0$ & $\mathbf{2.726-0}$ \\
$2$ & $8$ &  & $6.040-2$ & $6.556-2$ &  & $4.590-2$ & $4.852-2$ && $4.649-2$ & $4.879-2$ & $\mathbf{4.857-2}$ \\
$2$ & $9$ &  & $5.531-3$ & $8.699-3$ &  & $1.064-2$ & $1.114-2$ && $1.282-2$ & $1.298-2$ & $\mathbf{1.627-2}$ \\
$2$ & $10$ &  & $2.743-2$ & $4.177-2$ &  & $3.403-2$ & $4.580-2$&& $3.626-2$ & $4.684-2$ & $\mathbf{4.124-2}$ \\
$3$ & $4$ &  & $6.413-2$ & $7.493-2$ &  & $1.423-1$ & $1.436-1$ && $1.670-1$ & $1.716-1$ & $\mathbf{1.865-1}$ \\
$3$ & $5$ &  & $4.303-2$ & $4.516-2$ &  & $4.609-2$ & $4.633-2$ && $4.703-2$ & $4.814-2$ & $\mathbf{4.606-2}$ \\
$3$ & $6$ &  & $3.568-1$ & $4.775-1$ &  & $4.351-1$ & $5.344-1$ && $4.570-1$ & $5.450-1$ & $\mathbf{4.975-1}$ \\
$3$ & $7$ &  & $2.843-1$ & $3.113-1$ &  & $4.771-1$ & $4.829-1$ && $5.385-1$ & $5.496-1$ & $\mathbf{6.001-1}$ \\
$3$ & $8$ &  & $2.128-0$ & $2.145-0$ &  & $3.900-0$ & $3.860-0$ && $4.427-0$ & $4.458-0$ & $\mathbf{4.953-0}$ \\
$3$ & $9$ &  & $1.107-2$ & $1.745-2$ &  & $2.162-2$ & $2.239-2$ && $2.598-2$ & $2.608-2$ & $\mathbf{3.279-2}$ \\
$3$ & $10$ &  & $1.104-2$ & $1.412-2$ &  & $9.843-3$ & $1.038-2$&& $ 1.006-2$ & $1.050-2$ & $\mathbf{1.017-2}$\\
\hline
\end{tabular}
}
\end{center}
\vskip-6mm
\end{table}

\thanks{This work has been supported by the Tunisian research unit
05/UR/12- 04, the French one UMR 8112 and the bilateral
cooperation agreement between the Tunisian DGRS and the French
CNRS (project code 09/R 13-03, project No. 22637). The authors are
indebted to J. Dubau and M. Cornille for their invaluable help in
the use of the SST/DW/JAJOM computer codes.}

\References

\refb Aggarwal K. M. and Keenan F. P. 2004, Phys. Scr., 70, 222

\refb Dally K. G., Grant I. P., Johnson, C. T. et al. 1989,
Comput. Phys. Comm., 55, 424

\refb Eissner W., Jones M. and Nussbaumer H. 1974, Comput. Phys.
Comm., 8, 270

\refb Eissner W. 1998, Comput. Phys. Comm., 114, 295

\refb Elabidi H., Ben Nessib N., Cornille M. et al. 2008, J. Phys.
B.: Atom. Mol. Opt. Phys., 41, n° 025702

\refb Elabidi H., Sahal-Br\'{e}chot S. and Ben Nessib N. 2009,
EPJD, 54, 51

\refb Elabidi H. and Sahal-Br\'{e}chot S., 2011, EPJD, 61, 285

\refb Lozano J. M., Niimura M., Smith J. et al. 2001, Phys. Rev.
A, 63, 042713

\refb Saraph H. E. 1978, Comput. Phys. Comm., 15, 247
\end{document}